\documentclass[10pt,oneside,twocolumn,pra]{revtex4}

\usepackage{revsymb,graphicx}

\usepackage[english]{babel}

\begin{document}

\title{Weak Measurements with Entangled Probes}

\author{David Menzies and Natalia Korolkova}
\email{djm12@st-andrews.ac.uk} \affiliation{School of Physics and
Astronomy, University of St. Andrews, North Haugh, St. Andrews KY16
9SS, UK.}
\date{\today}

\begin{abstract}
Encoding the imaginary part of a weak value onto an initially
entangled probe can modify its entanglement content - provided the
probe observable can distinguish between states of different
entropies. Apart from fundamental interest, this result illustrates
the utility of the imaginary weak value as a calculational tool in
certain entanglement concentration protocols.
\end{abstract}

\maketitle

\section{Introduction}

Quantum mechanics has a reputation for being rather enigmatic. This
is largely due to the manner in which measurement is incorporated
into the theory. Despite the ongoing controversy over the
philosophical and interpretational issues surrounding quantum
measurement theory, it has nonetheless proved to be a fertile ground
for new exciting physics. This is particularly true for the case of
weak measurements as introduced in \cite{aharonov&vaidman:1990} by
{\it Aharonov} and {\it Vaidmann}. Such indirect quantum
measurements follow the template established by {\it Von Neumann}
\cite{vonneumman:1955} whose raison d'être is a circumvention of
strict requirement of a classical measuring device as advocated by
Bohr. Instead, the measurement device, or probe, is granted a
quantum description and the 'measurement' is regarded as the
entangling of the pointer degrees of freedom of the probe with the
eigen-states of the chosen observable of the system. To agree with
empirical observations, the probe is then postulated to collapse
into one of its pointer eigen-states by virtue of some intrinsic
mechanism possibly related to its macroscopic nature. Consequently,
the system collapses into a given eigen-state with its corresponding
eigen-value as the measurement result.

In contrast, the unique characteristics of a weak measurement
conspire to undermine the eigenvalue-measurement link axiom of
quantum mechanics. Firstly, the system of interest is both
pre-selected in $\vert i \rangle$ and post-selected in $\vert f
\rangle$. Secondly, the coupling strength governing the interaction
between system and probe is weak. As a result of these conditions, a
weak measurement does not encode the eigen-values on the probe.
Instead, it encodes the so-called weak value of the pre- and
post-selection:
\begin{equation}
O_W = \frac{\langle f \vert {\hat O} \vert i \rangle}{\langle f
\vert i \rangle}. \label{eqn1}
\end{equation}
Given its form as a normalized transition amplitude, it is obvious
that such weak values will not coincide with the eigen-values of the
observable (provided that $\vert i \rangle$ or $\vert f \rangle$ are
not eigenstates of $\hat O$). Perhaps more stunning, is the
realization that (\ref{eqn1}) can, in general, assume complex
values. The conclusion drawn from such weak measurements is that
observables between pre and post-selections can possess values
outside the eigen-value spectrum. In addition, one can `observe'
such values using a probe system coupled to the first via a weak
interaction.

Originally, {\it Aharonov} and {\it Vaidmann} restricted their
attention to interaction Hamiltonians of the form
\begin{equation}
\hat H_I = \kappa \hbar \hat O \otimes \hat P, \label{eqn2}
\end{equation}
where $\hat P$ is the conjugate pointer observable to $\hat Q$ with
$[\hat Q, \hat P] = i \hbar$. Moreover, they assumed that the probe
was initially prepared in a pure Gaussian distribution of pointer
eigen-states
\begin{equation}
\vert \Phi_i \rangle \propto \int dQ \; e^{-\frac{Q^2}{4 \Delta^2
Q}} \vert Q \rangle, \label{eqn3}
\end{equation}
with the final state of the probe given as
\cite{aharonov&vaidman:1990}:
\begin{equation}
\vert \Phi_f \rangle \propto (1 - i \kappa T O_W \hat P ) \vert
\Phi_i \rangle. \label{eqn4}
\end{equation}
In the intervening years, a number of authors
\cite{aharonav:2005,aharonov:2005,jozsa:2007} have characterize the
back-action of weak measurements on the probe in terms of the
moments of the probe's observables. In addition, other authors
\cite{johansen:2004,johansen&luis:2004} have illustrated that weak
values can be defined even when the system and probe exist in mixed
states.

In this article, we demonstrate that weak measurements can modify
certain non-classical features of the probe state. In particular we
show that the imaginary weak value has a r$\hat{\mbox{o}}$le in
modifying the entanglement content of the probe. However, this state
modification can only occur provided that the interaction
Hamiltonian fulfills a certain condition. Ultimately, the probe
observable must be able to distinguish between density matrices of
different entropies. The exact meaning of this will become clear in
section II. Apart from fundamental interest, we demonstrate, in
section III, the calculational utility of the weak value formalism
to aid in the understanding of certain entanglement concentration
protocols \cite{fiurasek:2003,menzies:2006}. When viewed in this
light, $\mbox{Im}(O_W)$ provides a selection criteria to constrain
the ingredients  $\{ \vert i \rangle, \hat O, \vert f \rangle, \hat
K \}$ and single out working examples of the protocol.

\section{Entangled Probes}

To set the scene, we assume a general configuration illustrated in
Figure 1.
\begin{figure}[!h] \centering
\includegraphics[height=40mm]{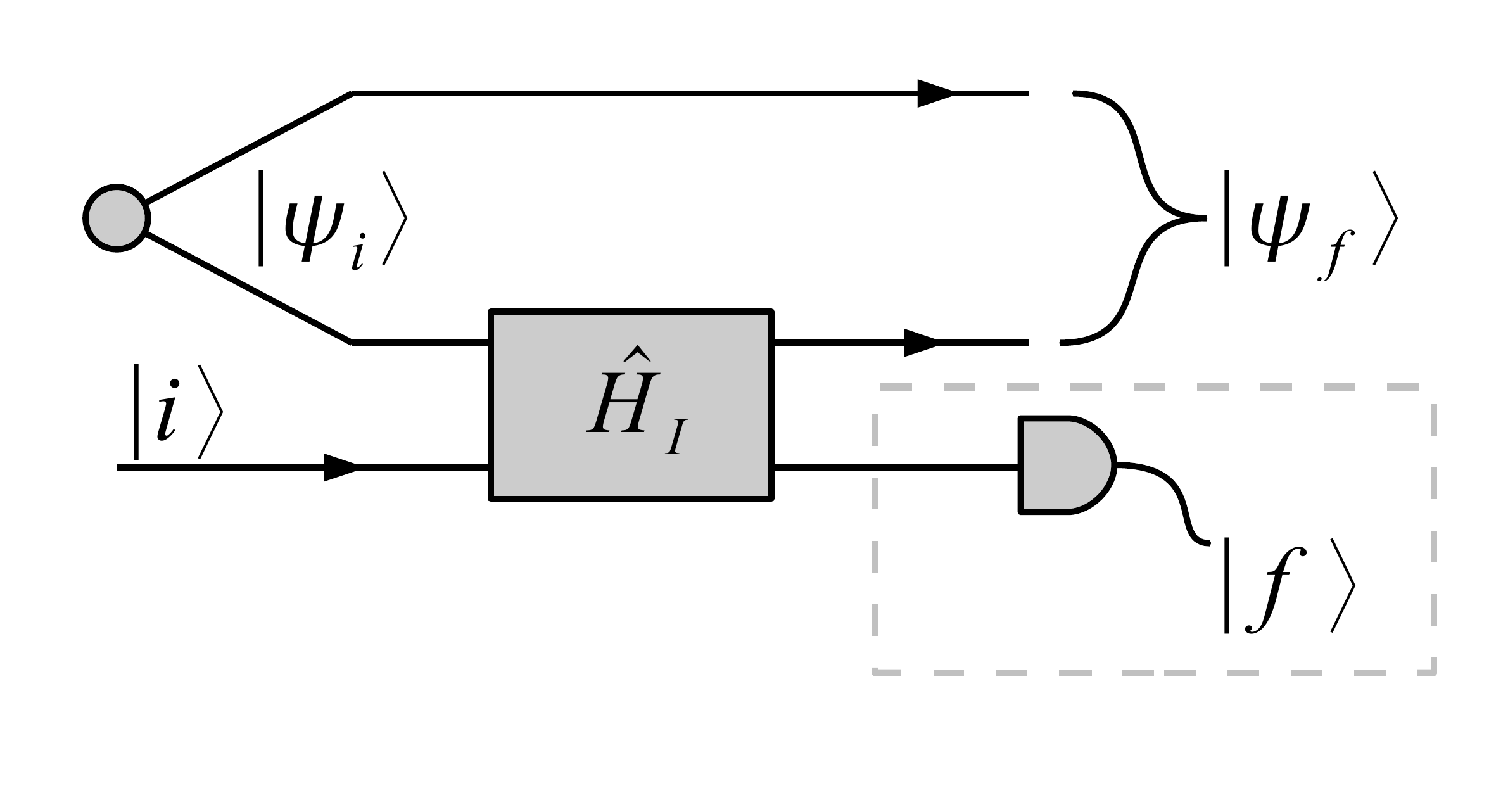}
\caption{\it Mixing one of the entangled sub-systems of $\vert
\psi_i \rangle$ with an ancilla system initially prepared in $\vert
i \rangle$, before post-selecting the ancilla in the state $\vert f
\rangle$.}
\end{figure}
The probe is initially prepared in an entangled state $\vert \psi_i
\rangle \in {\cal H}^{\otimes 2}$ with {\it Schmidt} decomposition:
\begin{equation}
\vert \psi_i \rangle = \sum_{k=1}^{K} s_k \vert a_k \rangle \vert
a_k \rangle. \label{eqn5}
\end{equation}
The usual properties are assumed with $\{ s_k \}_{k=1}^{K}$ obeying
$\sum_{k=1}^{K} s_k^2 = 1$, $s_k \geq 0\; \forall \: k \in [1,K]$
and $\langle a_k \vert a_j \rangle = \delta_{jk}$.  One subsystem of
the probe interacts with the system initially prepared in $\vert i
\rangle$. This mixing is provided by the interaction Hamiltonian
\begin{equation}
\hat H_I = \hbar \kappa \hat K \otimes \hat O, \label{eqn6}
\end{equation}
with, for simplicity, $\kappa = \mbox{const}$. In addition, it is
assumed that all of the systems have vanishing free Hamiltonians.
This can be done without loss of generality, provided we note that
all results are unique up to a suitable unitary transformation.
Furthermore, the observable $\hat K$ is required to admit the {\it
Schmidt} basis as an eigen-basis with $\hat K \vert a_k \rangle =
\lambda_k \vert a_k \rangle$. Consequently, the appropriate unitary
evolution operator generated by (\ref{eqn6}) is
\begin{equation}
\hat U_{BC} = \exp \left ( - \frac{i}{\hbar} \int_0^T dt \hat H_I
\right ) = e^{-i \varphi {\hat K} \otimes {\hat O} } \label{eqn7}
\end{equation}
and $\varphi = \kappa T$. Following the interaction, the observable
$\hat F$ is measured which results in the system being post-selected
in a particular eigenstate $\vert f \rangle$ and so the final probe
state is
\begin{equation}
\vert \psi_f \rangle \propto \langle f \vert \left ( \hat I \otimes
\hat U \right ) \vert \psi_i \rangle \vert i \rangle. \label{eqn8}
\end{equation}
The critical feature of weak measurements is the ``weakness'' of the
coupling between probe and system, thus it is assumed that $\varphi
<< 1$, meaning that only linear terms are kept:
\begin{equation}
\vert \psi_f \rangle \approx {\cal N} \left (  \langle f \vert i
\rangle \hat I - i \varphi \: \langle f \vert \hat O \vert i \rangle
( \hat I \otimes \hat K ) \right ) \vert \psi_i \rangle.
\label{eqn9}
\end{equation}
Introducing the weak value of $\hat O$ as
\begin{equation}
O_W = \frac{\langle f \vert \hat O \vert i \rangle}{\langle f \vert
i \rangle} \label{eqn10}
\end{equation}
and noting that $\tilde{\cal N} = {\cal N} {\langle f \vert i
\rangle}$ allows (\ref{eqn9}) to be given as
\begin{equation}
\vert \psi_f \rangle =\tilde{{\cal N}} \left ( \hat I_{AB} - i
\varphi \: O_W ( \hat I \otimes \hat K) \right ) \vert \psi_i
\rangle \label{eqn11},
\end{equation}
with normalisation constant $\tilde{\cal N}$
\begin{equation}
\tilde{\cal N}  \approx \frac{e^{i \phi}}{\sqrt{ 1 +2 \varphi \:
\mbox{Im}( O_W ) \langle \psi_i \vert ( \hat I \otimes \hat K )
\vert \psi_i \rangle }}, \label{eqn12}
\end{equation}
where $e^{i \phi}$ is an arbitrary global phase that can be set to
$\phi=0$ without loss of generality. Consequently, the final
entangled state is
\begin{eqnarray}
\vert \psi_f \rangle = \frac{\left ( \hat I - i \varphi \: O_W (
\hat  I \otimes \hat K) \right ) \vert \psi_i \rangle}{\sqrt{ 1 +2
\varphi \: \mbox{Im}( O_W ) \langle \psi_i \vert ( \hat I \otimes
\hat K ) \vert \psi_i \rangle }}. \label{eqn13}
\end{eqnarray}

From (\ref{eqn13}) it is clear that both the real and imaginary
parts of the weak value contribute towards the transformation of the
state. However, only the latter induces a non-unitary effect that is
responsible for the modification of the entanglement content of
(\ref{eqn5}). The verification of this effect requires the
demonstration of a quantitative change in the entanglement content
of the state. For bipartite pure states, the entanglement measure
derived from the {\it Von Neumann} entropy
\cite{bruss:2002,nielsen&chuang:2000}:
\begin{equation}
{\cal E}(\vert \Psi \rangle) = \{ {\cal S} \circ \mbox{Tr}_{j} \}
\vert \Psi \rangle \langle \Psi \vert = - \mbox{Tr}\left(
\hat{\rho}_j \ln \hat{\rho}_j \right ). \label{eqn14}
\end{equation}
The condition (\ref{eqn14}) holds for both subsystems $j=1,2$. The
symmetric nature of the {\it Schmidt} decomposition of (\ref{eqn5})
allows us to trace out any of the two subsystems with the same
result. The reduced density matrices are $\hat \sigma_i =
\mbox{Tr}_1(\hat \rho_i)$ and $\hat \sigma_f = \mbox{Tr}_1(\hat
\rho_f)$, where we denote $\hat \rho_i = \vert \psi_i \rangle
\langle \psi_i \vert$ and $\hat \rho_f = \vert \psi_f \rangle
\langle \psi_f \vert$. Accordingly, our starting point is the global
density matrix
\begin{equation}
\hat \rho_f = \frac{\hat \rho_i + 2\varphi \: \mbox{Im}(O_W) \hat{K}
\hat \rho_i}{1 + 2 \varphi \: \mbox{Im}( O_W ) \mbox{Tr}(\hat K \hat
\sigma_i)}, \label{eqn15}
\end{equation}
which leads to the reduced density matrix
\begin{equation}
\hat \sigma_f = \frac{\hat \sigma_i + 2\varphi \: \mbox{Im}(O_W)
\hat{K} \hat \sigma_i}{1 + 2 \varphi \: \mbox{Im}( O_W )
\mbox{Tr}(\hat K \hat \sigma_i)}. \label{eqn16}
\end{equation}
Moreover,
\begin{eqnarray}
\ln(\hat \sigma_f) = \ln(\hat \sigma_i + 2\varphi \: \mbox{Im}(O_W)
\hat{K} \hat \sigma_i) \nonumber \\ - \ln(1 + 2 \varphi \:
\mbox{Im}( O_W )
\mbox{Tr}(\hat K \hat \sigma_i)) \nonumber \\
\approx \ln(\hat \sigma) + 2 \varphi \: \mbox{Im}(O_W)  \delta
\hat{K}, \label{eqn17}
\end{eqnarray}
where $\delta \hat K = \hat K - \mbox{Tr}(\hat K \hat \sigma_i)$ and
the second line follows from keeping only linear terms in $\varphi$.
Hence
\begin{equation}
\hat \sigma_f \ln \hat \sigma_f = \frac{\hat \sigma_i \ln(\hat
\sigma_i) + 2 \varphi \: \mbox{Im}(O_W)( \delta \hat{K} \sigma_i +
\hat K \hat \sigma_i \ln \hat \sigma_i)}{1 + 2 \varphi \: \mbox{Im}(
O_W ) \mbox{Tr}(\hat K \hat \sigma_i)}, \label{eqn18}
\end{equation}
and so
\begin{equation}
{\cal S}(\hat \sigma_f) = \frac{{\cal S}(\hat \sigma_i) - 2 \varphi
\: \mbox{Tr}(\hat K \hat \sigma_i \ln \hat \sigma_i)}{1 + 2 \varphi
\: \mbox{Im}( O_W ) \mbox{Tr}(\hat K \hat \sigma_i)}. \label{eqn19}
\end{equation}
Finally, if we note that $\hat \omega =  \hat \sigma_i \ln \hat
\sigma_i / \mbox{Tr}(\hat \sigma_i \ln \hat \sigma_i)$ can be
formally identified as a density matrix in its own right, then
(\ref{eqn19}) becomes
\begin{equation}
\frac{{\cal S}(\hat \sigma_f)}{{\cal S}(\hat \sigma_i)} = \frac{1 +
2 \varphi \: \mbox{Im}( O_W ) \mbox{Tr}(\hat K \hat \omega)}{1 + 2
\varphi \: \mbox{Im}( O_W ) \mbox{Tr}(\hat K \hat \sigma_i)}.
\label{eqn20}
\end{equation}

The entanglement content of the probe is altered if and only if
${\cal S}_f \neq {\cal S}_i$ and so ${\cal S}_f/{\cal S}_i \neq 1$,
which is true if both $\mbox{Im}(O_W)$ and $\mbox{Tr}( \hat K ( \hat
\sigma_i -\hat \omega))$ cannot be zero. The requirement that
$\mbox{Im}(O_W) \neq 0$ is obvious from (\ref{eqn13}) as it
accompanies a non-unitary transformation of the probe state. This
follows from the properties of entanglement measures which are
designed to be non-increasing under local unitary operations
\cite{bruss:2002}.

On the other hand, the second simultaneous requirement that
$\mbox{Tr}( \hat K ( \hat \omega - \hat \sigma_i)) \neq 0$ is the
precise meaning to the claim that $\hat K$ must be able to
distinguish states of different entropies given earlier. This is
because the entropy of $\hat \sigma_i$ is identical to that of $\hat
\omega$ only when the global probe state is either separable or
maximally entangled. Essentially, the probe observable $\hat K_B$
must be able to witness the difference between the states $\hat
\sigma_i$ and $\omega$. It is instructive to compare this
requirement on $\hat K$ with the definition of an
entanglement-witness \cite{bruss:2002,vedral:2006} used in the
discussion of mixed state entanglement. Such a witness is a
self-adjoint operator $\hat W$ that can distinguish between the set
of separable states ${\cal S}$ and a particular entangled state
$\hat \rho_E$ via $\mbox{Tr}(\hat W \hat \rho) \geq 0 \: \forall
\hat \rho \in {\cal S}, \; \; \mbox{Tr}(\hat W \hat \rho_E) < 0$.
The essential difference between this and the r$\hat{o}$le played by
$\hat K$ is that the latter need only distinguish between two
states.

\section{Entanglement Concentration}

It is now widely acknowledged that the counter-intuitive features of
quantum states can also be interpreted as information theoretic
resources. This realization has provided ample motivation for the
study of quantum state engineering, with the aim of manufacturing,
enhancing or repairing the desired non-classical features of a
particular quantum state. Entanglement concentration protocols are
designed to augment the entanglement content of a particular
entangled state. From a state-engineering viewpoint, the weak
measurement with an entangled probe can be interpreted as a
entanglement distillation protocol. Essentially, by mixing a
subsystem with an ancilla which is both pre and post-selected can
augment the initial entanglement available in the global shared
state. In particular, this association can be a calculational aid
for {\it Procrustean} entanglement concentration protocols which
modify the {\it Schmidt} coefficients whilst preserving the basis:
\begin{equation}
\vert \psi_i \rangle = \sum_{k=1}^{d} s_k \vert k, k \rangle
\rightarrow \vert \psi_f \rangle = \sum_{k=1}^{d} t_k \vert k, k
\rangle. \label{eqn21}
\end{equation}
In our case, the output coefficients are a function of the weak
value of the ancilla $t_k = f(s_k,\varphi,O_W)$. When viewed in this
manner, we can use (\ref{eqn22}) to determine the requisite
conditions on $\{ \vert i \rangle, \hat O, \vert f \rangle, \hat K
\}$ that will collectively allow an entanglement concentration
effect. Entanglement concentration of the shared state (previously
known as the probe state) is given if ${\cal S}_f/{\cal S}_i > 1$
and hence
\begin{equation}
\mbox{Im}(O_W) \mbox{Tr}(\hat K (\hat \omega - \hat \sigma_i)) > 0,
\label{eqn22}
\end{equation}
Thus, the weak value formalism can be used in a quantum information
context to single out individual examples of entanglement
concentration protocols. Consequently, one can view $\mbox{Im}(O_W)$
as a calculational aid allowing one to pick out suitable ancilla
ingredients. Furthermore, in conjunction with condition on $\hat K$,
we find the required properties of the interaction Hamiltonian to
allow the desired effect.

The capacity of $\mbox{Im}(O_W)$ as a calculational tool in
entanglement concentration protocols has been previous noted by us
in \cite{menzies:2007} with the example of continuous-variable
states. The current work here can be regarded as a generalization to
arbitrary pure bipartite entangled states.

\section{Conclusion and Future Work}

In conclusion, we have demonstrate the consequences of employing an
initially entangled probe in weak measurements. In particular, we
have shown that the entanglement content of the probe can be
modified provided that the encoded weak value has a non-zero
imaginary component. In addition, it is critical that the probe
observable in the interaction Hamiltonian can distinguish between
states of different entropies. Thus, the probe observable plays a
r$\hat o$le similar to that of an entanglement witness with
$\mbox{Tr}(\hat K ( \hat \omega - \hat \sigma_i )) \neq 0$. Finally,
we presented evidence that the weak measurement model developed here
has applications in a family of entanglement concentration
protocols. Such protocols, illustrated in Figure 1, allowed for the
probabilistic enhancement of an initial entangled state following a
suitable interaction with an ancilla before sacrificing it to
measurement. In particular, the imaginary component of a weak value
can be used as selection criteria for the various ingredients $\{
\vert i \rangle, \hat O, \vert f \rangle, \hat K \}$  of the
ancilla.

The close connection between the imaginary weak value and
entanglement advocated here suggests that weak values could be
regarded as a resource that can be converted into entanglement. It
is then interesting to note that it has been previously established
that real weak values are a source of non-classicality
\cite{johansen&luis:2004,johansen:2004b}. A natural question then
arises: are imaginary weak values a source of non-classicality? If
true then it could be shown that imaginary weak values are capable
of modifying the non-classicality of the probe state. This would be
a further generalization to the result presented here. Obtaining an
answer to these questions is a future goal of our research.

\section*{Acknowledgements}

This work was supported by the {\it Engineering and Physical
Sciences Research Council} and by the EU project FP6-511004
COVAQIAL.

\bibliographystyle{unsrt}

\bibliography{entprobe}

\begin{thebibliography}{10}

\bibitem{aharonov&vaidman:1990}
Y.~Aharonov and L.~Vaidman.
\newblock {\em Phys. Rev. A. {\bf 41}, {\bf 1}, 11}, 1990.

\bibitem{vonneumman:1955}
J.~Von Neumann.
\newblock {\em Mathematical Foundations of Quantum Mechanics}.
\newblock Princeton Unversity Press, 1955.

\bibitem{aharonav:2005}
Y.~Aharonov and D.~Rohrlich.
\newblock {\em Quantum Paradoxes: Quantum Theory for the Perplexed}.
\newblock Wiley-Vch, 2006.

\bibitem{aharonov:2005}
Y.~Aharonov and A.~Botero.
\newblock {\em Phys. Rev. A. {\bf 72}, 052111}, 2005.

\bibitem{jozsa:2007}
R.~Jozsa.
\newblock {\em Phys. Rev. A. {\bf 76}, 0441031}, 2007.

\bibitem{johansen:2004}
L.~M. Johansen.
\newblock {\em Phys. Rev. Lett. {\bf 93}, {\bf 12}, 120402}, 2004.

\bibitem{johansen&luis:2004}
L.~M. Johansen and A.~Luis.
\newblock {\em Phys. Rev. A {\bf 70}, 052115}, 2004.

\bibitem{fiurasek:2003}
J.~Fiur$\grave{a}$$\check{s}$ek, L.~Mi$\check{s}$ta, and R.~Filip.
\newblock {\em Phys. Rev. A {\bf 67}, 022304}, 2003.

\bibitem{menzies:2006}
D.~Menzies and N.~V. Korolkova.
\newblock {\em Phys. Rev. A {\bf 74}}, page 042315, 2006.

\bibitem{bruss:2002}
D.~Bruss.
\newblock {\em Journ. Math. Phys. {\bf 43}}, page 4237, 2002.

\bibitem{nielsen&chuang:2000}
M.~Nielsen and I.~Chuang.
\newblock {\em Quantum Computation and Quantum Information}.
\newblock Cambridge University Press, 2000.

\bibitem{vedral:2006}
V.~Vedral.
\newblock {\em Introduction to Quantum Information Science}.
\newblock Oxford University Press, 2006.

\bibitem{menzies:2007}
D.~Menzies and N.~Korolkova.
\newblock {\em Phys. Rev. A {\bf 76}, 062310}, 2007.

\bibitem{johansen:2004b}
L.~M. Johansen.
\newblock {\em Phys. Lett. A. {\bf 329}, 184}, 2004.

\end{thebibliography}

\end{document}